# On the tulip flame formation: the effect of pressure waves

Chengeng Qian [a] and Mikhail A. Liberman [b] [*]

[a] *Aviation Key Laboratory of Science and Technology on High Speed and High Reynolds Number, Shenyang Key Laboratory of Computational Fluid Dynamics, Aerodynamic Force Research AVIC Aerodynamics Research Institute,*
*Shenyang 110034, China*

[b] *Nordita, KTH Royal Institute of Technology and Stockholm University, Hannes Alfvéns väg 12, 114 21 Stockholm, Sweden*

**Abstract**

The effects of pressure waves on the tulip flame (TF) formation in closed and semi-open tubes were studied using numerical simulations of the fully compressible Navier-Stokes equations coupled to a detailed chemical model for stoichiometric hydrogen–air mixture. Rarefaction waves generated by the decelerating flame are shown to be the primary physical process leading to the flame front inversion and the TF formation for spark or planar ignited flames in closed tubes. In the case of a spark ignited flame, the first rarefaction wave is generated by the flame, which is decelerating due to the reduction in flame surface area as the flame skirt comes in contact with the tube walls. Flame collisions with pressure waves in a closed tube result in additional deceleration stages and rarefaction waves that shorten the time of TF formation. The effect diminishes with increasing tube length (aspect ratio) because the number of flame collisions with reflected pressure waves is greater in a shorter tube, and the time of TF formation is maximum in semi-open tubes where there are no reflected pressure waves. The stages of flame deceleration in the case of planar ignition are due solely to the collision of the flame with the pressure waves reflected from the opposite closed end.

Chengeng Qian: qiancg@avic.com
Mikhail A. Liberman: mliber@nordita.org

# 1. Introduction

Flame propagation in closed or semi-open tubes is a fundamental chemical-physical model that provides a basic platform for the development of both analytical methods and numerical models important for many scientific and engineering combustion processes, e.g. explosions in confined spaces, power generation efficiency of internal combustion engines, pollutant emissions, etc. The early stages of combustion are of particular interest because they significantly influence the flame acceleration/deceleration that determine subsequent combustion regimes.

It is well known that the shape of a flame front ignited near the closed end of a tube and propagating toward the opposite closed or open end suddenly becomes "inverted" rapidly changing from a convex shape with the tip pointing forward to a concave shape with the tip pointing backward. This phenomenon, was first discovered experimentally by Ellis [1, 2] almost a century ago and subsequently named the "tulip flame" [3]. There is a large body of literature devoted to experimental, theoretical, and numerical studies performed to explain the physical origin (mechanism) of tulip flame formation. Different scenarios of tulip flame formation have been considered, including flame collision with a shock wave, Darrieus-Landau (DL) and Rayleigh-Taylor (RT) flame instabilities, flame-vortex interaction, etc. A long history of tulip flame research and related references can be found in the Dunn-Rankin [4] and Searby [5] reviews as well as in more recent work [6]. However, the physical processes that determine tulip flame formation remain among the least understood, including the role of pressure waves and the chemical reactivity of the combustible mixture. One of the reasons is that the meaning of tulip flame was not entirely definite and can be caused by various physical processes.

Many authors (see references in [6]) have suggested that flow characteristics behind the flame, such as vortices or backflow, are responsible for flame front inversion and tulip flame formation. On the contrary, in [6] it was shown that rarefaction waves are the primary physical process leading to flame front inversion and tulip flame formation. It is known [7] that when the piston begins to move out of the tube with negative acceleration, it generates a simple rarefaction wave whose head propagates forward ($x > 0$) with a sound speed through the gas at rest, and a reveres flow is established between the head of rarefaction wave and the piston, where velocity is negative everywhere, with maximum (negative) magnitude at the piston surface, and decreases monotonically in magnitude toward the head of rarefaction wave. A flame during the deceleration stage acts like a piston moving out of the tube. But in this case the resulting unburned gas flow is a superposition of the flow created by the rarefaction wave and the flow created by the flame during prior acceleration stages. As a result, the velocity of the unburned gas decreases, especially near the flame front, and the boundary layer thickness in the unburned flow near the flame front increases. This means that the velocity profile of the unburned gas close to the flame front acquires a tulip shape: the velocity is minimal at the tube axis, increases towards the wall, reaches the maximum value, then decreases in the boundary layer and disappears at the tube wall. In the thin flame model [8, 9], the velocity of each small part of the flame front is equal to the sum of the laminar flame velocity with which flame propagates relative to the unburned gas and the velocity of the unburned gas immediately ahead of that part of the flame front with which this part is entrained by the unburned flow. Therefore, the flame front replicates (mirrors) the shape of the velocity profile in the unburned gas, taking on a tulip shape.

In this paper we investigate the role of pressure waves in the formation of tulip flames. What causes the flame inversion, how it depends on the aspect ratio of a tube and the ignition source (spark or planar) used to ignite the flame, and what is the role of pressure waves. In the following sections, we present the results of a series of two- and three-dimensional numerical simulations of flame propagation in closed tubes and in semi-open tubes. It should be emphasized that described above mechanism of flame front inversion and tulip flame formation is a purely hydrodynamic process that proceeds faster than the development of flame inherent instabilities. For example, the time of the DL instability development estimated as the inverse increment is $t_{DL} \approx D / 2\pi U_f \Theta^{1/2}$ the hydrodynamic timescale is $t_{HD} \sim D / a_s$, and the condition $t_{HD} < t_{DL}$ reduces to inequality

$$U_f / a_s < 1 / 2\pi\Theta^{1/2}. \qquad (1)$$

For a hydrogen/air flame this inequality is $2\pi\Theta^{1/2} U_f / a_s \approx 0,09 \ll 1$, where $U_f = 2.1 m/s$ is the laminar flame velocity, $a_s = 408,77 m/s$ is sound speed, $\Theta = \rho_u / \rho_b = 7.8$ is expansion coefficient, and $D$ is a channel width. This means that we consider the tulip flame formation as a purely hydrodynamic process, in agreement with the results of recent experimental studies by Ponizy et al [10], whose conclusion is that the tulip flame formation is a purely hydrodynamic process not associated with any flame front instabilities.

## 2. Numerical models and physical parameters

*2.1 Two-dimensional DNS*

The two-dimensional computational domains, that were modeled, are the rectangular channels of width $D = 1 cm$ and aspect ratios of $L/D = 6$, with both ends closed and a rectangular channel of width $D = 1 cm$ with an open right end. The two-dimensional, time-dependent, reactive compressible Navier-Stokes equations including molecular diffusion, thermal conduction and viscosity are solved using a fifth-order WENO finite-difference method [11]. We used a detailed chemical kinetics model for a stoichiometric hydrogen–air mixture consisting of 19 reactions and 9 species [12]. The resolution and convergence tests (grid independence) similar to those in previous publications [13, 14] were thoroughly performed to ensure that the resolution is adequate to capture details of the problem in question and to avoid computational artifacts. The input parameters describing a stoichiometric mixture of hydrogen/air are: initial pressure $P_0 = 1\,atm$, initial temperature $T_0 = 298\,K$, initial density $\rho_0 = 8.5 \cdot 10^{-4}\,g/cm^3$, laminar flame velocity $U_f = 2.1\,m/s$, laminar flame thickness $L_f = 0.0325\,cm$, adiabatic flame temperature $T_b = 2350\,K$, expansion coefficient $\Theta = \rho_u/\rho_b = 7.8$, specific heat ratio $\gamma = C_P/C_V = 1.399$, sound speed $a_s = 408.77\,m/s$.

*2.2 Three-dimensional LES*

Three-dimensional case was modeling using Large Eddy Simulation (LES) and a one-step chemical model calibrated to reproduce velocity-pressure dependaence as in a detailed chemical model. The thickened flame model [15] was used, when the flame front is thickened artificially without changing the laminar flame velocity. For a premixed laminar flame, the laminar flame velocity $U_f$ and the flame thickness $L_f$ can be expressed as

$$U_f \propto (D_{th} A)^{1/2}, \quad L_f \propto \frac{D_{th}}{U_f} = (D_{th}/A)^{1/2}, \tag{2}$$

where $D_{th}$ and $A$ are thermal diffusivity and pre-exponential factor. Supposing that $D_{th}$ and $A$ are replaced by $FD_{th}$ and $A/F$, it is easy to check that $L_f$ increases by a factor $F$ while $U_f$ remains unchanged. A comparison of the flame dynamics obtained with DNS for a flame propagating in a two-dimensional channel with the flame dynamics obtained using LES have shown that qualitatively the results of DNS and LES are quite close. More detailes about simulations, resolution and convergence tests are in [6, 14].

## 3. Results

*3.1 Two-dimensional channel*

First consider the dynamics of a flame ignited by a spark of small size compared to the width of the tube $D = 1 cm$ at the left end of a two-dimensional tube with a relatively small aspect ratio $L / D = 6$ and both ends closed. Figure 1 shows computed schlieren images and streamlines at selected times during a tulip flame formation.

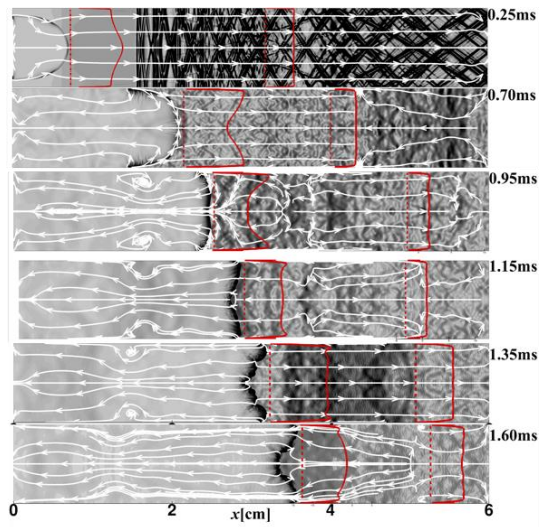

Fig. 1 Sequences of computed schlieren images, stream lines and the unburned gas velocity profiles.

It can be seen that the convex at the beginning flame front up to 0.95 ms is inverted into a concave tulip-shaped one within a short time ≈ 0.1-0.15ms, which is shorter than the time required for the development of the DL instability for a perturbation of wavelength $\lambda = D / 2$. The streamlines in Fig. 1 show reverse flows in the unburned gas, which is a consequence of the rarefaction waves. The oblique pressure waves seen in Fig.1 were generated by the convex flame during the acceleration phases, and they propagated down the tube, repeatedly colliding and reflecting off the side walls, reflecting off the right end of the tube, and running back and forth along the tube.

As mentioned earlier, the velocity in the unburned gas decreases and the boundary layer thickness increases in the vicinity of the flame front, resulting in a tulip-shaped velocity profile in the unburned gas near the flame front. The profiles of the velocities in the unburned gas are shown in Fig.1 near the flame and far from the flame. The local velocity of each point at the flame front in the laboratory reference frame is $\vec{U}_{fL} = \vec{U}_f + \vec{u}_+(x,y)$, which means that the shape of the flame front "replicates" the shape of the unburned flow velocity profile. Figure 2(a) shows the time evolution of the unburned gas velocity at the tube axis $u_+(y=0)$ and at the "inner edge" of the boundary layer $u_+(y=0.40)$ calculated at 0.5 mm ahead of the flame front. Figure 2(b) shows the time evolution of pressure $P$, the flame front velocities at $y=0$ and $y=0.40\,\text{cm}$.

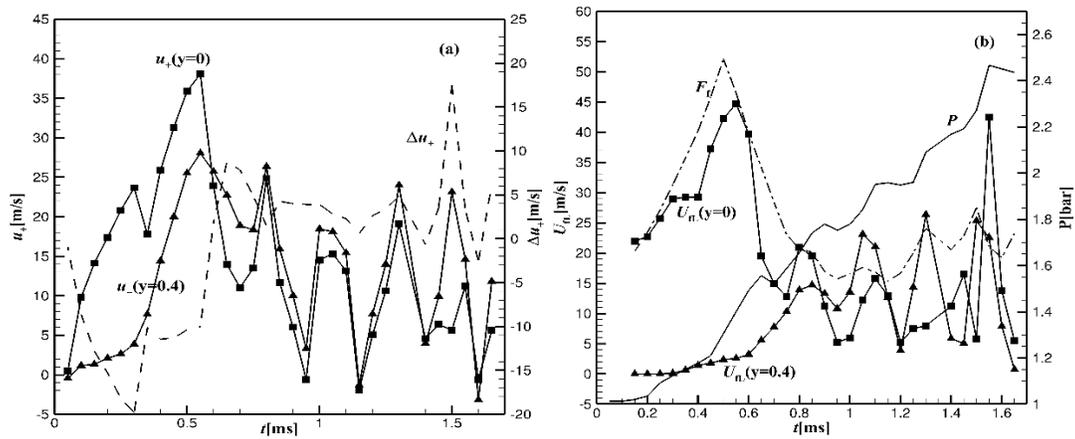

**Fig. 2: a)** Time evolution of the unreacted flow velocities $u_+(X_f + 0.5mm, y)$ at 0.5 mm ahead of the flame front at $y=0$, $y=0.4\,cm$, and the difference $\Delta u_+ = u_+(y=0.4) - u_+(y=0)$; **b)** Time evolution of the flame surface area, the flame front velocities at $y=0$, $y=0.4\,cm$, pressure and the difference . $\Delta U_{Fl} = U_{Fl}(y=0.4) - U_{Fl}(y=0)$.

It can be seen that each collision of the flame with the pressure wave reflected from the right end of the tube is associated with an additional phase of flame deceleration and rarefaction wave, increasing the magnitudes of $\Delta u_+ = u_+(y=0.4) - u_+(y=0)$ and $\Delta U_{Fl} = U_{Fl}(y=0.4) - U_{Fl}(y=0)$. However, the intensity of the first rarefaction wave associated with the reduction of the flame surface area at 0.55 ms, which can be characterized by the magnitude of the flame deceleration, is greater than the intensity of the rarefaction waves caused by flame collisions with pressure waves.

The intensity of the first rarefaction wave is sufficient to create a tulip flame in a semi-open tube when there are no reflected pressure waves. Figure 3 shows computed schlieren images and streamlines at selected times during a tulip flame formation for a flame ignited by a spark at the left closed end of the tube of width $D = 1\,cm$ and propagating to the opposite open end.

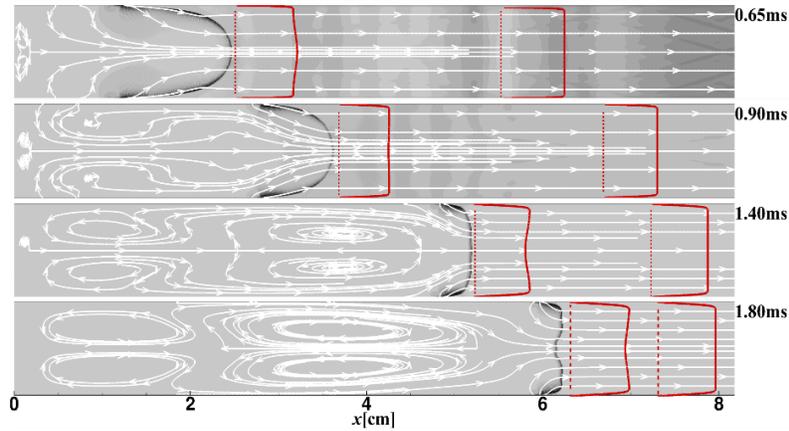

Fig. 3. Sequences of computed schlieren images, stream lines and the unburned gas velocity profiles for the premixed hydrogen–air flame propagating in a half-open tube.

Figure 4(a, b) shows the time evolution of the unburned gas velocity at the tube axis $y = 0$ and at the inner edge of the boundary layer $y = 0.4\,cm$ calculated at 0.5 mm ahead of the flame front and the velocities of the flame front at these points. The way the flame moves and takes a tulip shape in a semi-open tube is pretty similar to what happens in a closed tube, but it takes longer time since there are no reflected pressure waves.

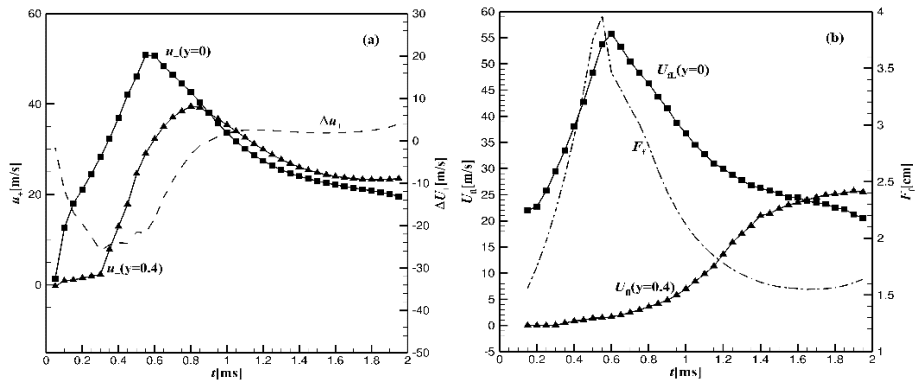

**Fig. 4: a)** Evolution of the unburned gas velocity at the tube axis and at the inner edge of the boundary layer ($y = 0.4\,cm$) calculated at 0.5 mm ahead of the flame front; **b)** time evolution of the flame front velocities at $y = 0$ and $y = 0.4\,cm$.

The dynamics of a flame ignited by a planar, narrow band of high-temperature combustion products occupying the entire width of the tube (planar ignition) is quite different. Figure 5(a) shows the time evolution of the unburned gas velocities at $y=0$ and $y=0.42\,cm$ at the distance of 0.5mm ahead of the flame front and the increase of $\Delta u_+ = u_+(y = 0.42) - u_+(0)$ leading to the formation of a tulip flame. Fig. 5b shows the temporal evolution of the flame surface area $F_f$ and the local flame front velocities at the tube axis $y = 0$ and near the wall $y = 0.42\,cm$, for a flame initiated by planar ignition at the left closed end of a tube of width $D = 1\,cm$, aspect ratio $L/D = 6$ and the closed opposite end. In this case, unlike a flame ignited by a small spark, the flame skirt touches the side walls of the tube from the beginning. Therefore, the flame surface area $F_f$ increases monotonically as the lateral edges of the flame skirt extend backward along the tube walls. The flame tip velocity $U_{fL}(y=0)$ and the local flame front velocity near the wall, $U_{fL}(y = 0.42\,cm)$ oscillate as the result of the flame collisions with pressure waves reflected from the right closed end of the tube. Nevertheless, it can be seen in Fig.5b, that after each oscillation the local flame front velocity near the wall, $U_{fL}(y = 0.42\,cm)$ exceeds the flame tip velocity, $U_{fL}(y = 0)$. Finally, after several collisions $U_{fL}(y = 0.42\,cm)$ significantly exceeds $U_{fL}(y = 0)$. This tendency is a consequence of several weak rarefaction waves resulting from the flame deceleration due to the flame collisions with the pressure wave, which leads to an increase in the velocity of the edges of the flame front near the wall compared to the flame front velocity at the tube axis.

Fig. 6 shows a sequence of numerical schlieren images, streamlines and the unburned gas velocity profiles for selected times during tulip flame formation for a flame initiated by planar ignition in a tube of width $D = 1 cm$, $L / D = 6$.

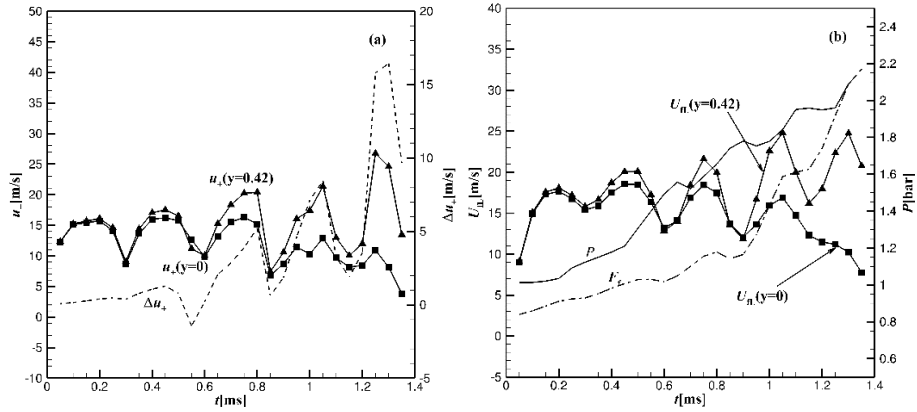

**Fig. 5. a)** Time evolution of the unburned flow velocities 0.5mm ahead of the flame front at the tube axis $y = 0$ and at near the wall $y = 0.42 cm$; **b)** Time evolution of the flame surface area, and local velocities of the flame front at the tube axis $y = 0$ and at near the wall $y = 0.42 cm$.

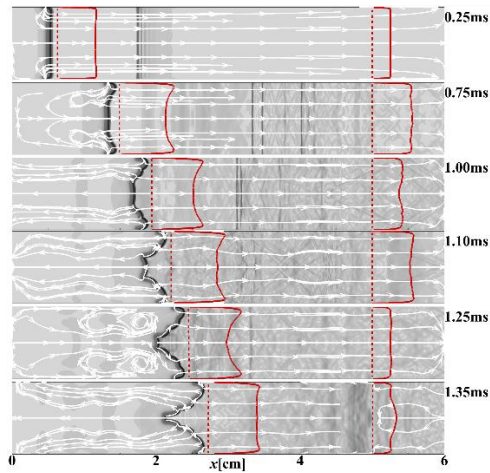

Fig. 6. Time sequence of computed Schlieren images and streamlines for the planar ignition of the flame propagating in the tube with both ends closed.

*3.2 Three-dimensional channel*

Flame dynamics and tulip flame formation in a 3D rectangular channel is qualitatively similar to this in a 2D channel. The most significant difference is the higher rate of flame acceleration and deceleration by increasing or decreasing the flame surface area [9], so that the maximum flame speed achieved in the acceleration phase is almost twice as high in the 3D case as in the 2D case.

Figure 7(a, b) shows: a) the evolution of unburned flow velocities at 0.5mm ahead of the flame front at the cross section $(x, y, z = 0)$, at $y = 0$, $y = 0.3 cm$ and $\Delta u_+ = u_+(y = 0.3 cm) - u_+(y = 0)$, for the flame propagating in the rectangular channel of length $L = 6 cm$ and cross section $D \times D = 1 cm^2$; b) the time evolution of the flame surface area $F_f$, the flame front velocities at the tube axis $U_{fl}(y = 0)$ and near the walls $U_{fl}(y = 0.3 cm)$, and pressure $P$.

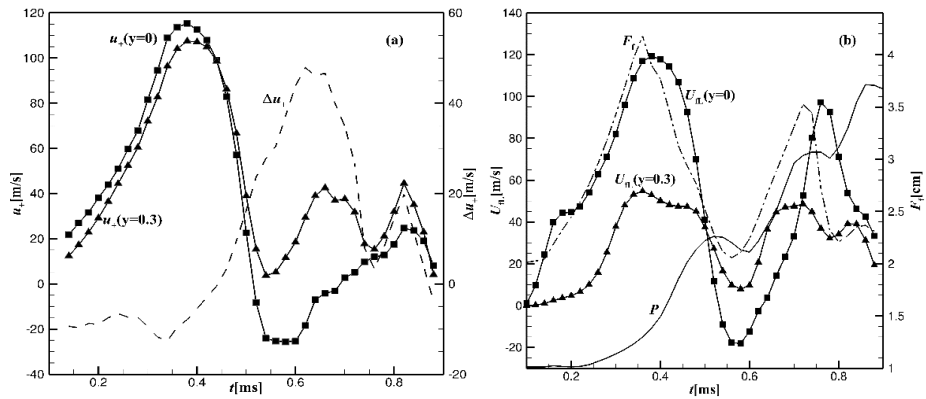

**Fig. 7. a)** Time evolution of the unburned flow velocities 0.5mm ahead of the flame front at the tube axis $y = z = 0$ and at near the wall $y = 0.3 cm$; **b)** Time evolution of the flame surface area, and local velocities of the flame front at the tube axis and at near the wall. Flame propagates in the tube with both ends closed, $L/D = 6$, $D \times D = 1 cm^2$.

Figure 8 shows a sequence of numerical schlieren images, stream lines and velocity profiles in the unburned gas in cross sections $(x, y, z = 0)$, which locations is shown by the dashed lines.

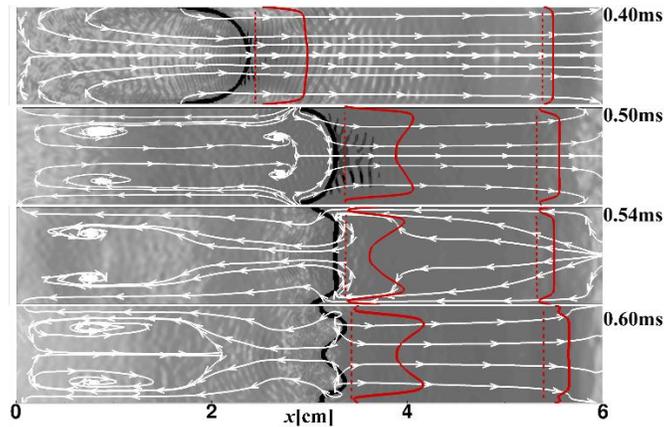

**Fig. 8.** Sequence of schlieren images, stream lines and velocity profiles in the unburned gas for the flame propagating in the 3D tube with both closed ends $L = 6\text{cm}$, cross section $1 \times 1\text{cm}^2$, shown is cross section $(x, y, z = 0)$.

## 4. Conclusions

The paper shows that the inversion of the flame front from convex to concave shape and the formation of the tulip flame is a purely hydrodynamic process, not related to the instabilities of the flame front, and therefore occurs faster than the characteristic times of the instabilities inherent to the flame, in agreement with experimental results [10]. Rarefaction waves are shown to be the primary physical process leading to flame front inversion and tulip flame formation for spark or planar ignition sources in closed tubes. Another important conclusion is that the inversion of the flame front and the formation of a tulip flame occur much faster for 3D flames than for 2D flames. Although the results of two-dimensional modeling are qualitatively quite similar to those observed in the experiment, they should be interpreted with caution because there are significant quantitative differences between two-dimensional and three-dimensional modeling.